\documentclass[aoas,MSNbibl,nameyear,dvips]{arximspdf}

\doi{10.1214/13-AOAS708} 
\volume{8}
\issue{1}
\pubyear{2014}
\firstpage{577}
\lastpage{594}

\makeatletter
\newcommand{\rrvert}{\vert}
\newcommand{\llvert}{\vert}
\newcommand{\overset}{\stackrel}
\newtheorem{theorem}{Theorem}
\newproclaim{remark}{Remark}
\makeatother

\begin{document}
\begin{frontmatter}

\title{Predictive regressions for macroeconomic data}
\runtitle{Predictive regressions}

\begin{aug}
\author[A]{\fnms{Fukang} \snm{Zhu}\thanksref{t1,m1}\ead[label=e1]{zfk8010@163.com}},
\author[B]{\fnms{Zongwu} \snm{Cai}\thanksref{t2,m2}\ead[label=e2]{caiz@ku.edu}}
\and
\author[C]{\fnms{Liang} \snm{Peng}\corref{}\thanksref{t3,m3}\ead[label=e3]{peng@math.gatech.edu}}
\runauthor{F. Zhu, Z. Cai and L. Peng}
\affiliation{Jilin University\thanksmark{m1}, University of Kansas and
Xiamen University\thanksmark{m2},\\ and Georgia Institute of
Technology\thanksmark{m3}}
\address[A]{F. Zhu\\
School of Mathematics\\
Jilin University\\
Changchun, Jilin 130012\\
China\\
\printead{e1}} 
\address[B]{Z. Cai\\
Department of Economics\\
University of Kansas\\
Lawrence, Kansas 66045\\
USA\\
and\\
Wang Yanan Institute for Studies in Economics\\
Fujian Key Laboratory of Statistical Sciences\\
Xiamen University\\
Xiamen 361005\\
China\\
\printead{e2}}
\address[C]{L. Peng\\
School of Mathematics\\
Georgia Institute of Technology\\
Atlanta, Georgia 30332\\
USA\\
\printead{e3}}
\end{aug}
\thankstext{t1}{Supported in part by National Natural Science
Foundation of China Grants Nos. 11371168, 11001105 and 11271155,
Specialized Research Fund for the Doctoral Program of Higher Education
No.~20110061110003,
Science and Technology Developing Plan of Jilin Province No.~20130522102JH and 985 Project of Jilin University.}
\thankstext{t2}{Supported in part by National Natural Science
Foundation of China Grants No. 71131008 (Key Project) and No. 70971113.}
\thankstext{t3}{Supported in part by NSF Grant DMS-10-05336.}

\received{\smonth{6} \syear{2013}}
\revised{\smonth{9} \syear{2013}}

%
\begin{abstract}
Researchers have constantly asked whether stock returns can be
predicted by some macroeconomic data. However, it is known that
macroeconomic data may exhibit nonstationarity and/or heavy tails,
which complicates existing testing procedures for predictability. In
this paper we propose novel empirical likelihood methods based on some
weighted score equations to test whether the monthly CRSP value-weighted
index can be predicted by the log dividend-price ratio or the log
earnings-price ratio. The new methods work well both theoretically and
empirically regardless of the predicting variables being stationary or
nonstationary or having an infinite variance.
\end{abstract}

%
\begin{keyword}
\kwd{Autoregressive process}
\kwd{empirical likelihood}
\kwd{long memory process}
\kwd{nearly integrated}
\kwd{predictive regressions}
\kwd{unit root}
\kwd{weighted estimation}
\end{keyword}

\end{frontmatter}

\section{Introduction}\label{sec1}
It is well documented in the literature that predictive regression
models have been widely used in economics and finance for the
evaluation of the mutual fund performance, the optimization of the
asset allocations, the conditional capital asset pricing and others.
In particular, it is used to check the predictability of asset returns
by various lagged financial and economic variables, such as the log
dividend-price ratio, the log earnings-price ratio, the log
book-to-market ratio, the dividend yield, the term spread and default
premium, the interest rates as well as other financial and state
economic variables.

Our motivation for this research is trying to answer the question in
the financial econometrics literature on whether the monthly CRSP
(Center for Research in Security Prices) value-weighted index can be
predicted by using the macroeconomic data such as the log
dividend-price ratio or the log earnings-price ratio as well as other
economic data like interest rates.
To answer this question, we need a statistical model. By following the
convention in the financial econometrics literature, we use the
following simple predictive regression model which
assumes that observations $\{(X_t,Y_t)\}_{t=1}^n$ follow the following
structural model:
%
\begin{equation}
\label{Model} \cases{ Y_t=\alpha+\beta X_{t-1}+U_t,
\vspace*{2pt}\cr
X_t=\theta+\phi X_{t-1}+V_t}
\end{equation}
with $X_0$ being a constant. Here, $Y_t$ denotes a predictable
variable, say, the asset return like the CRSP value-weighted index,
$X_t$ denotes a predicting variable, such as financial instruments like
the log dividend-price ratio or the log earnings-price ratio, and
$(U_1,V_1), \ldots, (U_n,V_n)$ are
independent and identically distributed (i.i.d.) innovations with zero
means but $U_t$ and $V_t$ might be correlated.
Our main purpose of this study is to examine the existence of the
predictability of asset returns by some
financial variables such as the log dividend-price ratio or the log
earnings-price ratio. To achieve our goal, we need to construct a
confidence interval for $\beta$ in (\ref{Model}) or to test the null
hypothesis of no predictability ($H_0\dvtx \beta=0$). The detailed report of
analyzing the aforementioned real example is given in Section~\ref{sec4}.

The empirical literature on the predictability of asset returns is
rather large. In particular, estimating $\beta$ and testing the null
hypothesis of no predictability $H_0\dvtx \beta=0$
are receiving much attention in the recent literature of financial
econometrics. For example, \citet{r21} showed that the least
squares estimator for $\beta$ based on the first equation in (\ref
{Model}) is biased in finite sample since the estimation procedure
ignores the dependence between $U_t$ and $V_t$. Since then, several
bias-corrected estimation procedures and corresponding hypothesis tests
have been proposed in the literature when the sequence $\{X_t\}$ is
stationary (i.e., $|\phi|< 1$) and/or integrated/nearly integrated
(i.e., $\phi=1-\gamma_\phi/n$ for some $\gamma_\phi\ge0$).
Some references include but are not limited to
\citet{r1}, \citet{r5}, \citet{r9}, \citet{r14}, \citet{r15}, \citet{r2}, \citet{r3} and the references therein.

By assuming that the joint distribution of the two innovations $(U_t,
V_t)$ in (\ref{Model}) is a bivariate normal, \citet{r5}
proposed a new Bonferroni \mbox{$Q$-}test,
based on the infeasible uniform most powerful test, and showed that
this new test is more powerful than the Bonferroni $t$-test of
\citet{r6} in the sense of Pitman efficiency.
However, the normality assumption might not be satisfied for real
applications and the implementation of the Bonferroni \mbox{$Q$-}test can be
somewhat complicated, because it requires searching several tables as in
\citet{r4}, which depend heavily on both the
Dickey--Fuller generalized least squares (DF-GLS) statistic and $\delta$
being the correlation coefficient between $U_t$ and $V_t$. Moreover,
the theoretical
justification of the Bonferroni \mbox{$Q$-}test given in \citet{r5} heavily depends on the assumptions of known covariance of
innovations, known shifts in the model and that the predicting variable
is nonstationary
and has a finite variance.
It remains unjustified when these unknown quantities are replaced by
some estimators and/or the predicting variable is stationary or has an
infinite variance.

Now, the question is how to construct a confidence interval for $\beta$
or to test whether $\beta$ equals a given value, say, zero, without
knowing that the predicting variable is stationary or nonstationary or
has an infinite variance. Obviously, none of those methods mentioned
above work since the asymptotic limit of any one of them depends on
whether the predicting variable is stationary or nonstationary or has
an infinite variance. Moreover, it is impossible to distinguish these
cases without imposing further model assumptions. To illustrate this
difficulty, let us look at the simple least squares estimator of $\beta
$ in (\ref{Model}), given by
\[
\hat\beta_{\mathrm{LSE}}=\frac{n\sum_{t=1}^nY_tX_{t-1}-(\sum_{t=1}^nY_t)(
\sum_{t=1}^nX_{t-1})}{n\sum_{t=1}^nX_{t-1}^2-(\sum_{t=1}^nX_{t-1})^2}.
\]
Clearly, $\hat\beta_{\mathrm{LSE}}$ can be re-expressed as follows:
\[
\hat\beta_{\mathrm{LSE}}-\beta=\frac{n\sum_{t=1}^nU_tX_{t-1}-(\sum_{t=1}^nU_t)(
\sum_{t=1}^nX_{t-1})}{n\sum_{t=1}^nX_{t-1}^2-(\sum_{t=1}^nX_{t-1})^2}.
\]
It is known that $n^{-1}\sum_{t=1}^nX_{t-1}$ and $n^{-1}\sum_{t=1}^nX_{t-1}^2$ do not converge in probability to some constants
when the AR(1) process $\{X_t\}$ is integrated/nearly integrated.
Therefore, the asymptotic limit of $\hat\beta_{\mathrm{LSE}}$ is totally
different for the stationary and nonstationary cases; see \citet{r5} and \citet{r3}.
On the other hand, when $\{X_t\}$ and $\{U_t\}$ are two independent
random samples with heavy tails,
\citet{r20} derived the asymptotic limit of $\hat\beta
_{\mathrm{LSE}}$, which is very complicated too. Therefore, if one wants to
construct a confidence interval for $\beta$ or to test $H_0\dvtx \beta=\beta
_0$ for a given value $\beta_0$ based on the asymptotic limit of $\hat
\beta_{\mathrm{LSE}}$, one has to distinguish the case between stationarity and
nonstationarity, and between finite variance and infinite variance.
This seems infeasible in the real implementation.
Moreover, even if one can distinguish these cases, it is still a
difficult task to obtain critical points by directly estimating or
simulating the asymptotic limit when the sequence $\{X_t\}$ is
integrated/nearly integrated and/or has an infinite variance.
As an alternative way, a bootstrap method may be employed to obtain
critical values.
However, it is well known in the literature that the full sample
bootstrap method is inconsistent for a nearly integrated or infinite
variance AR process.
Instead, one has to employ the subsample bootstrap method and face the
difficulty of choosing the subsample size; see \citet{r13} and
\citet{r10} for details.

To overcome the aforementioned difficulties and problems, in this
paper, by applying the empirical likelihood method to some weighted
score equations, we propose new methods to construct a confidence
interval for $\beta$ or to test $H_0\dvtx  \beta=\beta_0$ without
distinguishing whether the sequence $\{X_t\}$ is stationary or
nonstationary (integrated or nearly integrated) or has an infinite variance.
As a powerful nonparametric likelihood approach, empirical likelihood
method has been extended and applied to many different settings
including time series models since Owen (\citeyear{r17}, \citeyear{r18}) introduced the
method. See \citet{r19} for an overview.

The rest of this paper is organized as follows. Section~\ref{sec2} is devoted
to presenting the methodologies and some asymptotic results. A
simulation study is reported in Section~\ref{sec3}, which shows the good finite
sample performance of the new methods. The detailed analysis of the
monthly CRSP value-weighted index
is reported in Section~\ref{sec4} to highlight the practical usefulness of the
proposed methods. Section~\ref{sec5} concludes the paper. All theoretical proofs
are relegated to Section~\ref{sec6}.

\section{Methodology and asymptotic properties}\label{sec2}
First, we consider that observations $\{(X_t,Y_t)\}$ follow the model
%
\begin{equation}
\label{Mod1} \cases{ Y_t=\beta X_{t-1}+U_t,
\vspace*{2pt}\cr
X_t=\theta+\phi X_{t-1}+e_t,
\vspace*{2pt}\cr
B(L)e_t=V_t,}
\end{equation}
where $L^ie_t=e_{t-i}$, $B(L)=1-(\sum_{i=1}^pb_iL^i)$, $B(1)\neq0$,
all the roots of $B(L)$ are fixed and less than one in absolute value,
and $(U_1,V_1),\ldots,(U_n,V_n)$ are i.i.d. random vectors with zero means.

As shown in \citet{r8}, the empirical likelihood method
fails for nonstationary AR processes in the sense that Wilks' theorem
does not hold.
It~is also known that the asymptotic limit of the least squares
estimator for $\phi$ in the second equation of (\ref{Mod1}) is a stable
law rather than a normal distribution when $e_t$~has an infinite
variance. Hence, it is expected that Wilks' theorem fails for a direct
application of the empirical likelihood method to the score equation
via the first equation in (\ref{Mod1}) when the sequence $\{X_t\}$ is
either nonstationary or has an infinite variance.

Recently, \citet{r16} proposed minimizing the weighted least squares\break
$\sum_{t=1}^n\{X_t-\theta-\phi X_{t-1}\}^2w(X_{t-1})$ for some weight
function $w(\cdot)$ so as to ensure a normal limit whenever $e_t=V_t$
has a finite or infinite variance.
\citet{r7} combined the weighted idea with the empirical
likelihood method to construct a confidence interval for $\phi$
whenever the sequence $\{X_t\}$ is stationary or nearly integrated, but
has a finite variance. Here, we propose using the weighted idea
together with the empirical likelihood method to construct a confidence
interval for $\beta$ rather than $\phi$ regardless of the sequence $\{
X_t\}$ being stationary or nearly integrated or having an infinite variance.
More specifically, we define the empirical likelihood function for
$\beta$ as
%
\begin{equation}
\label{emplik0} \qquad L_n(\beta)=\sup \Biggl\{\prod
_{t=1}^n (np_t)\dvtx  p_1\ge0,
\ldots,p_n\ge0, \sum_{t=1}^np_t=1,
\sum_{t=1}^np_t
Z_t(\beta)=0 \Biggr\},
\end{equation}
where $Z_t(\beta)=(Y_t-\beta X_{t-1})X_{t-1}/\sqrt{1+X_{t-1}^2}$.
It follows from the Lagrange multiplier technique that
\[
l_n(\beta)=-2\log L_n(\beta)=2\sum
_{t=1}^n\log \bigl\{1+\lambda Z_t(\beta)
\bigr\},
\]
where $\lambda=\lambda(\beta)$ satisfies
\[
\sum_{t=1}^n\frac{ Z_t(\beta)}{1+\lambda Z_t(\beta)}=0.
\]
The following theorem shows that Wilks'k theorem holds for the above
proposed empirical likelihood method.

\begin{theorem}
\label{th1}
Suppose model (\ref{Mod1}) holds with either $|\phi|<1$ or $\phi
=1-\gamma_\phi/n$ for some $\gamma_\phi\ge0$. Furthermore, we assume
that $EU_1=0, E|U_1|^{2+q}<\infty$ for some $q>0$,
and the distribution of $V_t$ is in the domain of attraction of a
stable law with index $\alpha^\ast\in(0, 2]$. Then, $l_n(\beta_0)$
converges in distribution to a chi-square limit with one degree of
freedom as $n\to\infty$, where $\beta_0$ denotes the true value of
$\beta$.
\end{theorem}

\begin{remark}
If $EV_t^2<\infty$, then the distribution of $V_t$ is in the domain of
attraction of a stable law with index $\alpha^\ast=2$.
When the distribution of $V_t$ is in the domain of attraction of a
stable law with index $\alpha^\ast=2$, $EV_t^2$ may be infinite, but
$E|V_t|^{\gamma^\ast}<\infty$ for any $\gamma^\ast\in(0, 2)$.
When the distribution of $V_t$ is in the domain of attraction of a
stable law with index $\alpha^\ast\in(0, 2)$, we have $E|V_t|^{\gamma
^\ast}<\infty$
for $\gamma^\ast<\alpha^\ast$ and $E|V_t|^{\gamma^\ast}=\infty$ for
$\gamma^\ast>\alpha^\ast$. The reader is referred to \citet{r11} for
details on stable laws.
\end{remark}

Next, we consider a more general model than (\ref{Mod1}) by including
an intercept for~$Y_t$:
%
\begin{equation}
\label{Mod2} \cases{ Y_t=\alpha+\beta X_{t-1}+U_t,
\vspace*{2pt}\cr
X_t=\theta+\phi X_{t-1}+e_t,
\vspace*{2pt}\cr
B(L)e_t=V_t,}
\end{equation}
where $L^ie_t=e_{t-i}$, $B(L)=1-(\sum_{i=1}^pb_iL^i)$, $B(1)\neq0$,
all the roots of $B(L)$ are fixed and less than one in absolute value,
and $(U_1,V_1),\ldots,(U_n,V_n)$ are i.i.d. random vectors. Once again,
our observations are $\{(X_t,Y_t)\}_{t=1}^n$.

As before, one may apply the empirical likelihood method to the
following estimating equations:
\[
\sum_{t=1}^n(Y_t-\alpha-\beta
X_{t-1})=0
\]
and
\[
\sum_{t=1}^n(Y_t-
\alpha-\beta X_{t-1})X_{t-1}/\sqrt{1+X_{t-1}^2}=0.
\]
It is clear that when $\{X_t\}$ is integrated/nearly integrated,
$n^{-1}\sum_{t=1}^nU_tX_{t-1}/\break \sqrt {1+X_{t-1}^2}$ does not converge in probability to a constant. Instead,
it converges in distribution. Therefore, the joint limit of $\frac{1}{\sqrt n}\sum_{t=1}^n(Y_t-\alpha_0-\beta_0 X_{t-1})$ and $\frac{1}{\sqrt n}\sum_{t=1}^n(Y_t-\alpha_0-\beta_0X_{t-1})X_{t-1}/\sqrt {1+X_{t-1}^2}$ is no longer a bivariate normal distribution.
Hence, Wilks' theorem for the above empirical likelihood method fails
when $\{X_t\}$ is nonstationary, which
is due to the intercept $\alpha$.

To overcome the above difficulty, one may employ the difference method
to get rid of $\alpha$ by using $Y_{t+1}-Y_t$. In such a case, the
sequence $\{X_{t+1}-X_t\}_{t=1}^n$ becomes stationary when $\phi=1$.
Therefore, inferences for $\beta$ based on the differences become much
less efficient with rate $\sqrt n$ instead of $n$ when the sequence $\{
X_t\}_{t=1}^n$ is nonstationary.
Another issue on applying the empirical likelihood method based on the
difference $Y_{t+1}-Y_t$ is that the new errors $\{U_{t+1}-U_t\}
_{t=1}^n$ are not independent any more.
Here, we propose to split the sample into two parts and then to use the
differences with a very large lag to get rid of the intercept before
applying the empirical likelihood method. More specifically,
put $m=[n/2]$, $\tilde Y_t=Y_t-Y_{t+m}$, $\tilde X_t=X_t-X_{t+m}$, and
$\tilde U_t=U_t-U_{t+m}$ for $t=1,\ldots,m$. Then, we have
\[
\tilde Y_t=\beta\tilde X_{t-1}+\tilde U_t\qquad
\mbox{for } t=1,\ldots,m.
\]
Based on the above equation, we define the empirical likelihood
function for $\beta$ as
%
\begin{equation}\label{emlik}
\tilde L_n(\beta)=\sup \Biggl\{\prod_{t=1}^m(mp_t)\dvtx
p_1\ge0,\ldots, p_m\ge0, \sum
_{t=1}^mp_t=1, \sum
_{t=1}^mp_t\tilde Z_t(
\beta)=0 \Biggr\},\hspace*{-20pt}
\end{equation}
where $\tilde Z_t(\beta)=(\tilde Y_t-\beta\tilde X_{t-1})\tilde
X_{t-1}/\sqrt{1+\tilde X_{t-1}^2}$. By the Lagrange multiplier
technique, we have
\[
\tilde l_n(\beta)=-2\log\tilde L_n(\beta)=2\sum
_{t=1}^m\log \bigl\{1+\tilde \lambda\tilde
Z_t(\beta) \bigr\},
\]
where $\tilde\lambda=\tilde\lambda(\beta)$ satisfies
\[
\sum_{t=1}^m\frac{\tilde Z_t(\beta)}{1+\tilde\lambda\tilde Z_t(\beta)}=0.
\]
The following theorem shows that Wilks' theorem holds for the above
proposed empirical likelihood method.
%
\begin{theorem}
\label{th2}
Under conditions of Theorem~\ref{th1}, $\tilde l_n(\beta_0)$ converges
in distribution to a chi-square distribution with
one degree of freedom as $n\to\infty$, where $\beta_0$~denotes the
true value of $\beta$.
\end{theorem}

Based on the above theorems, an empirical likelihood confidence
interval for $\beta_0$ with level $b$ can be obtained as
\[
I_b= \bigl\{\beta\dvtx  l_n(\beta)\le\chi_{1,b}^2
\bigr\}\quad\mbox{and}\quad\tilde I_{b}= \bigl\{\beta\dvtx  \tilde
l_n(\beta)\le\chi_{1,b}^2 \bigr\}
\]
for models (\ref{Mod1}) and (\ref{Mod2}), respectively,
where $\chi^2_{1,b}$ denotes the $b$th quantile of a chi-square
distribution with one degree of freedom.
Therefore, the implementation for constructing the confidence interval
is straightforward without estimating any additional quantities.
Indeed, the function ``\textit{emplik}'' in the R package [see \citet{r22}]
can be employed to compute $l_n(\beta)$ and $\tilde l_n(\beta) $ as
easily as we do in the simulation study below.

\section{A Monte Carlo simulation study}\label{sec3}

In this section we investigate the finite sample behavior of the
proposed empirical likelihood methods for testing \mbox{$H_0\dvtx  \beta=0$}
against $H_a\dvtx  \beta\neq0$.
We compare our new methods with the bootstrap method and the
Bonferroni \mbox{$Q$-}test proposed in \citet{r5} in terms of
both size and power.

First, we calculate the rejection region based on the least squares
estimator $\hat\beta_{\mathrm{LSE}}$ by using the bootstrap
method to obtain critical points. More specifically, we first estimate
$\alpha, \beta, \theta, \phi, b_j's$ in (\ref{Mod2}) by\vspace*{-2pt} least squares
estimators, which results in an estimator for $(U_t, V_t)$, say, $(\hat
U_t, \hat V_t)$. Next, we draw $1000$ random samples with size $n-1$
from $(\hat U_t, \hat V_t)$, say, $(U_t^{*(j)}, V_t^{*(j)})$ for
$t=1,\ldots,n-1$ and $j=1,\ldots,1000$. Using model (\ref{Mod2}) with
estimated $\alpha, \beta, \theta, \phi, b_j's$, we obtain the
bootstrap samples $\{(X_t^{*(j)}, Y_t^{*(j)})\}_{t=1}^{n-1}$.
For\vspace*{1pt} each $j$, we use the bootstrap sample $X_{1}^{*(j)},\ldots,X_{n-2}^{*(j)},Y_{2}^{*(j)},\ldots,Y_{n-1}^{*(j)}$ to estimate
$\beta$ by the least squares approach again. Therefore, the rejection
region can be obtained
based on these $1000$ bootstrapped least squares estimators for $\beta
$. Note that such a bootstrap method is theoretically inconsistent when
the sequence $\{X_t\}$ is either nearly integrated or has an infinite variance.

Next, we implement the Bonferroni \mbox{$Q$-}test given in \citet{r5}. Note that the theoretical derivation of the tests in \citet{r5} assumes that $\alpha, \theta$ and the covariance of
$(U_t, V_t)$ are known and $\phi$ is near one although the
implementation of the Bonferroni \mbox{$Q$-}test given in \citet{r4} has no such requirements. Theoretically, one may suspect that
the Bonferroni \mbox{$Q$-}test is inconsistent when $\alpha$ and $\theta$ are
replaced by their corresponding estimators and $\phi$ is not close to
one. In order to validate this conjecture, we compute the Bonferroni
\mbox{$Q$-}test by using both the true values and the estimated values of
$\alpha$ and $\theta$.
Since the implementation of the Bonferroni \mbox{$Q$-}test requires to search
several tables in \citet{r4}, which depend on both the
DF-GLS statistic and $\delta$ being the correlation coefficient between
$U_t$ and $V_t$, and are only designed for constructing a 90\%
two-sided confidence interval or 95\% one-sided confidence interval, we
fix $\delta=-0.75$ in the model setup. That is,
we consider model (\ref{Mod2}) with $U_t\sim N(0,1)$, $\varepsilon_t\sim
t(\nu)$, $\delta=-0.75$, $V_t=\delta U_t + \frac{\sqrt{1-\delta
^2}}{\sqrt{\nu/(\nu-2)}}\varepsilon_t$ if $\nu>2$ and $V_t=\delta
U_t+\varepsilon_t$ if $\nu\le2$, where $U_1,\ldots,U_n$ and $\varepsilon
_1,\ldots,\varepsilon_n$ are two independent random samples.
We also choose $\alpha=0$, $\beta=a/\sqrt{n}$, $\theta=0$, $\phi=
0.9$,\vadjust{\goodbreak} $0.99$,~$1$, $p=1$, $b_1=0$, $-0.5$, $\nu=4$, $1.5$, $0.5$ and
repeat 10,000 times with sample size $n=100$ and $300$ from the above
setting. Hence, results for $a=0$ correspond to the size.

\begin{table}
\tabcolsep=0pt
\caption{Empirical sizes are reported for testing $H_0\dvtx  \beta=0$
against $H_a\dvtx  \beta\neq0$ with level 10\% for the proposed empirical
likelihood test in (\protect\ref{emplik0}) with known $\alpha$ (EL1),
the proposed empirical likelihood test in (\protect\ref{emlik}) with
unknown $\alpha$ (EL2), the normal approximation based on bootstrap
method (NA), the Bonferroni \mbox{$Q$-}test in \citet{r5} with
known $\alpha$ and $\theta$ (BQ1), and the Bonferroni \mbox{$Q$-}test with
unknown $\alpha$ and $\theta$ (BQ2). Sample size $n=100$}\label{t1}
\begin{tabular*}{\tablewidth}{@{\extracolsep{\fill}}@{}lccccc@{}}
\hline
$\bolds{(a,\phi, \nu, b_1)}$&\textbf{EL1}&\textbf{EL2}&\textbf{NA}&\textbf{BQ1}&\textbf{BQ2}\\
\hline
$(0,0.9,4,0)$ & 0.1019 & 0.1084 & 0.0971 & 0.0460 & 0.0297\\
$(0,0.99,4,0)$ & 0.1022 & 0.0976 & 0.0682 & 0.0955 & 0.0341\\
$(0,1,4,0)$ & 0.1048 & 0.1155 & 0.0666 & 0.1085 & 0.0363\\
$(0,0.9,1.5,0)$ & 0.1005 & 0.1039 & 0.0458 & 0.0613 & 0.0418\\
$(0,0.99,1.5,0)$ & 0.1036 & 0.1053 & 0.0358 & 0.1576 & 0.0445\\
$(0,1,1.5,0)$ & 0.1038 & 0.1077 & 0.0417 & 0.1881 & 0.0449\\
$(0,0.9,0.5,0)$ & 0.1007 & 0.1028 & 0.0125 & 0.0864 & 0.0634\\
$(0,0.99,0.5,0)$ & 0.1025 & 0.1035 & 0.0168 & 0.2124 & 0.0600\\
$(0,1,0.5,0)$ & 0.1035 & 0.1031 & 0.0283 & 0.2406 & 0.0630\\
$(0,0.9,4,-0.5)$ & 0.1052 & 0.1070 & 0.1088 & 0.0688 & 0.0156\\
$(0,0.99,4,-0.5)$ & 0.1042 & 0.0963 & 0.0817 & 0.2326 & 0.0137\\
$(0,1,4,-0.5)$ & 0.1055 & 0.1172 & 0.0750 & 0.2505 & 0.0126\\
$(0,0.9,1.5,-0.5)$ & 0.1008 & 0.1000 & 0.0500 & 0.0593 & 0.0354\\
$(0,0.99,1.5,-0.5)$ & 0.1036 & 0.1051 & 0.0377 & 0.1809 & 0.0427\\
$(0,1,1.5,-0.5)$ & 0.1053 & 0.1065 & 0.0414 & 0.2078 & 0.0440\\
$(0,0.9,0.5,-0.5)$ & 0.0978 & 0.1009 & 0.0130 & 0.0902 & 0.0639\\
$(0,0.99,0.5,-0.5)$ & 0.1041 & 0.1030 & 0.0159 & 0.2200 & 0.0568\\
$(0,1,0.5,-0.5)$ & 0.1018 & 0.1023 & 0.0248 & 0.2532 & 0.0700\\
\hline
\end{tabular*}
\end{table}

\begin{table}[t]
\tabcolsep=0pt
\caption{Empirical sizes are reported for testing $H_0\dvtx  \beta=0$
against $H_a\dvtx  \beta\neq0$ with level 10\% for the proposed empirical
likelihood test in (\protect\ref{emplik0}) with known $\alpha$ (EL1),
the proposed empirical likelihood test in (\protect\ref{emlik}) with
unknown $\alpha$ (EL2), the normal approximation based on bootstrap
method (NA), the Bonferroni \mbox{$Q$-}test in \citet{r5} with
known $\alpha$ and $\theta$ (BQ1), and the Bonferroni \mbox{$Q$-}test with
unknown $\alpha$ and $\theta$ (BQ2). Sample size $n=300$}\label{t2}
\begin{tabular*}{\tablewidth}{@{\extracolsep{\fill}}@{}lccccc@{}}
\hline
$\bolds{(a,\phi, \nu, b_1)}$&\textbf{EL1}&\textbf{EL2}&\textbf{NA}&\textbf{BQ1}&\textbf{BQ2}\\
\hline
$(0,0.9,4,0)$ & 0.1036 & 0.1048 & 0.1061 & 0.0251 & 0.0213\\
$(0,0.99,4,0)$ & 0.1035 & 0.0869 & 0.0752 & 0.0636 & 0.0335\\
$(0,1,4,0)$ & 0.1063 & 0.1051 & 0.0627 & 0.0936 & 0.0311\\
$(0,0.9,1.5,0)$ & 0.1058 & 0.1087 & 0.0580 & 0.0424 & 0.0383\\
$(0,0.99,1.5,0)$ & 0.1004 & 0.1055 & 0.0362 & 0.1105 & 0.0438\\
$(0,1,1.5,0)$ & 0.0980 & 0.1072 & 0.0391 & 0.1914 & 0.0424\\
$(0,0.9,0.5,0)$ & 0.1005 & 0.1031 & 0.0081 & 0.0617 & 0.0551\\
$(0,0.99,0.5,0)$ & 0.0966 & 0.1012 & 0.0079 & 0.1454 & 0.0565\\
$(0,1,0.5,0)$ & 0.0970 & 0.0989 & 0.0194 & 0.2334 & 0.0552\\
$(0,0.9,4,-0.5)$ & 0.1052 & 0.1070 & 0.1088 & 0.0688 & 0.0156\\
$(0,0.99,4,-0.5)$ & 0.1043 & 0.0885 & 0.0790 & 0.2358 & 0.0193\\
$(0,1,4,-0.5)$ & 0.1084 & 0.1071 & 0.0656 & 0.3239 & 0.0208\\
$(0,0.9,1.5,-0.5)$ & 0.1032 & 0.1076 & 0.0573 & 0.0443 & 0.0374\\
$(0,0.99,1.5,-0.5)$ & 0.1010 & 0.1034 & 0.0392 & 0.1288 & 0.0442\\
$(0,1,1.5,-0.5)$ & 0.0975 & 0.1049 & 0.0378 & 0.2104 & 0.0430\\
$(0,0.9,0.5,-0.5)$ & 0.1044 & 0.1031 & 0.0082 & 0.0610 & 0.0534\\
$(0,0.99,0.5,-0.5)$ & 0.0958 & 0.1020 & 0.0082 & 0.1454 & 0.0566\\
$(0,1,0.5,-0.5)$ & 0.0963 & 0.0994 & 0.0192 & 0.2363 & 0.0558\\
\hline
\end{tabular*}
\end{table}

We also calculate the empirical likelihood functions in both (\ref
{emplik0}) and (\ref{emlik}) by using the R package ``\textit{emplik}'' in
\citet{r22}, that is, we consider both known and unknown $\alpha$.
In Tables \ref{t1}~and~\ref{t2}, we report the sizes for these tests.
From these two tables, we observe that the proposed empirical
likelihood methods have a size close to the nominal level 0.1 whenever
the sequence $\{X_t\}$ is stationary or near-integrated or has an
infinite variance.
The normal approximation method via the bootstrap method only works for
the case of $(\phi,\nu)=(0.9,4)$, that is, it fails when the sequence
$\{X_t\}$ is either nearly integrated or has an infinite variance.
This is not surprising because this empirical evidence is in line with
the theory provided by \citet{r10}, \citet{r13}.
Furthermore, it is interesting to see that the Bonferroni \mbox{$Q$-}test
seems to be only working for the case of $(\phi,\nu,b_1)=(1,4,0)$ with
known $\alpha$ and $\theta$. Therefore, it remains cautious to
employ the Bonferroni \mbox{$Q$-}test in \citet{r5}
due to the complicated implementation and lack of theoretical justification.

\begin{table}
\caption{Empirical powers are reported for testing $H_0\dvtx  \beta=0$
against $H_a\dvtx  \beta\neq0$ with level 10\% for the proposed empirical
likelihood test in (\protect\ref{emplik0}) with known $\alpha$ (EL1),
the proposed empirical likelihood test in (\protect\ref{emlik}) with
unknown $\alpha$ (EL2), the normal approximation based on bootstrap
method (NA), the Bonferroni \mbox{$Q$-}test in \citet{r5} with
known $\alpha$ and $\theta$ (BQ1), and the Bonferroni \mbox{$Q$-}test with
unknown $\alpha$ and $\theta$ (BQ2). Sample size $n=100$}\label{t3}
\begin{tabular*}{\tablewidth}{@{\extracolsep{\fill}}@{}lccccc@{}}
\hline
$\bolds{(a,\phi, \nu, b_1)}$&\textbf{EL1}&\textbf{EL2}&\textbf{NA}&\textbf{BQ1}&\textbf{BQ2}\\
\hline
$(-0.3,0.9,4,0)$ & 0.1831 & 0.1582 & 0.2097 & 0.0556 & 0.0428\\
$(-0.3,0.99,4,0)$ & 0.3872 & 0.1578 & 0.2167 & 0.1750 & 0.0908\\
$(-0.3,1,4,0)$ & 0.4613 & 0.1721 & 0.2312 & 0.2566 & 0.1412\\
$(-0.1,0.9,1.5,0)$ & 0.2366 & 0.1957 & 0.1529 & 0.1957 & 0.1670\\
$(-0.1,0.99,1.5,0)$ & 0.5168 & 0.3175 & 0.2770 & 0.4347 & 0.2865\\
$(-0.1,1,1.5,0)$ & 0.5991 & 0.3478 & 0.3495 & 0.5159 & 0.3448\\
$(-0.002,0.9,0.5,0)$ & 0.6002 & 0.5814 & 0.4149 & 0.6869 & 0.6673\\
$(-0.002,0.99,0.5,0)$ & 0.7925 & 0.7176 & 0.6168 & 0.8113 & 0.7461\\
$(-0.002,1,0.5,0)$ & 0.8215 & 0.7235 & 0.6740 & 0.8370 & 0.7679\\
$(-0.3,0.9,4,-0.5)$ & 0.1495 & 0.1348 & 0.1694 & 0.1348 & 0.0498\\
$(-0.3,0.99,4,-0.5)$ & 0.2834 & 0.1170 & 0.1512 & 0.3804 & 0.0569\\
$(-0.3,1,4,-0.5)$ & 0.3518 & 0.1214 & 0.1541 & 0.4551 & 0.0769\\
$(-0.1,0.9,1.5,-0.5)$ & 0.1798 & 0.1589 & 0.1027 & 0.1679 & 0.1281\\
$(-0.1,0.99,1.5,-0.5)$ & 0.3907 & 0.2330 & 0.1707 & 0.3990 & 0.1986\\
$(-0.1,1,1.5,-0.5)$ & 0.4683 & 0.2538 & 0.2180 & 0.4733 & 0.2366\\
$(-0.002,0.9,0.5,-0.5)$ & 0.5337 & 0.5149 & 0.3513 & 0.6321 & 0.6108\\
$(-0.002,0.99,0.5,-0.5)$ & 0.7350 & 0.6536 & 0.5297 & 0.7627 & 0.6801\\
$(-0.002,1,0.5,-0.5)$ & 0.7648 & 0.6634 & 0.5903 & 0.7921 & 0.7072\\
\hline
\end{tabular*}
\end{table}

\begin{table}
\tabcolsep=0pt
\caption{Empirical powers are reported for testing $H_0\dvtx  \beta=0$
against $H_a\dvtx  \beta\neq0$ with level 10\% for the proposed empirical
likelihood test in (\protect\ref{emplik0}) with known $\alpha$ (EL1),
the proposed empirical likelihood test in (\protect\ref{emlik}) with
unknown $\alpha$ (EL2), the normal approximation based on bootstrap
method (NA), the Bonferroni \mbox{$Q$-}test in \citet{r5} with
known $\alpha$ and $\theta$ (BQ1), and the Bonferroni \mbox{$Q$-}test with
unknown $\alpha$ and $\theta$ (BQ2). Sample size $n=300$}\label{t4}
\begin{tabular*}{\tablewidth}{@{\extracolsep{\fill}}@{}lccccc@{}}
\hline
$\bolds{(a,\phi, \nu, b_1)}$&\textbf{EL1}&\textbf{EL2}&\textbf{NA}&\textbf{BQ1}&\textbf{BQ2}\\
\hline
$(-0.3,0.9,4,0)$ & 0.1787 & 0.1461 & 0.2182 & 0.0587 & 0.0541\\
$(-0.3,0.99,4,0)$ & 0.4674 & 0.2742 & 0.3933 & 0.2788 & 0.1953\\
$(-0.3,1,4,0)$ & 0.6547 & 0.3258 & 0.4457 & 0.5156 & 0.3933\\
$(-0.1,0.9,1.5,0)$ & 0.2318 & 0.1955 & 0.2367 & 0.2457 & 0.2337\\
$(-0.1,0.99,1.5,0)$ & 0.6937 & 0.5122 & 0.5717 & 0.6361 & 0.5411\\
$(-0.1,1,1.5,0)$ & 0.8484 & 0.6328 & 0.7032 & 0.7972 & 0.6867\\
$(-0.002,0.9,0.5,0)$ & 0.8285 & 0.8153 & 0.8367 & 0.9240 & 0.9204\\
$(-0.002,0.99,0.5,0)$ & 0.9730 & 0.9568 & 0.9495 & 0.9794 & 0.9717\\
$(-0.002,1,0.5,0)$ & 0.9870 & 0.9698 & 0.9738 & 0.9898 & 0.9829\\
$(-0.3,0.9,4,-0.5)$ & 0.1437 & 0.1262 & 0.1728 & 0.1651 & 0.1026\\
$(-0.3,0.99,4,-0.5)$ & 0.3314 & 0.1895 & 0.2636 & 0.4296 & 0.1054\\
$(-0.3,1,4,-0.5)$ & 0.5156 & 0.1992 & 0.2779 & 0.6398 & 0.2266\\
$(-0.1,0.9,1.5,-0.5)$ & 0.1695 & 0.1555 & 0.1535 & 0.2030 & 0.1836\\
$(-0.1,0.99,1.5,-0.5)$ & 0.5276 & 0.3659 & 0.3593 & 0.5354 & 0.3705\\
$(-0.1,1,1.5,-0.5)$ & 0.7258 & 0.4767 & 0.5041 & 0.7153 & 0.5116\\
$(-0.002,0.9,0.5,-0.5)$ & 0.7595 & 0.7492 & 0.7562 & 0.8866 & 0.8831\\
$(-0.002,0.99,0.5,-0.5)$ & 0.9510 & 0.9278 & 0.9080 & 0.9611 & 0.9488\\
$(-0.002,1,0.5,-0.5)$ & 0.9764 & 0.9472 & 0.9487 & 0.9783 & 0.9649\\
\hline
\end{tabular*}
\end{table}

\begin{table}
\tabcolsep=0pt
\caption{Empirical sizes and powers are reported for testing $H_0\dvtx
\beta=0$ against $H_a\dvtx  \beta\neq0$ with level 10\% for the proposed
empirical likelihood test in (\protect\ref{emplik0}) with known $\alpha
$ (EL1) and the proposed empirical likelihood test in (\protect\ref
{emlik}) with unknown $\alpha$ (EL2), where the general weight function
$w(t)=(1+|t|^h)^{1/h}$ is employed. Sample size $n=300$}\label{t5}
\begin{tabular*}{\tablewidth}{@{\extracolsep{\fill}}@{}lcccccc@{}}
\hline
 & \multicolumn{3}{c}{\textbf{EL1}}& \multicolumn{3}{c@{}}{\textbf{EL2}}\\[-6pt]
 & \multicolumn{3}{c}{\hrulefill}& \multicolumn{3}{c@{}}{\hrulefill}\\
$\bolds{(a,\phi, \nu, b_1)}$ & $\bolds{h=1}$&$\bolds{h=2}$&$\bolds{h=4}$&$\bolds{h=1}$&$\bolds{h=2}$&$\bolds{h=4}$\\
\hline
$(0,0.9,4,0)$ & 0.0970&0.0963&0.0956&0.1005&0.1006&0.0989\\
$(0,0.99,4,0)$ & 0.1025&0.1008&0.1006&0.0868&0.0853&0.0833\\
$(0,1,4,0)$ & 0.1097&0.1067&0.1053&0.1099&0.1043&0.1022\\
$(0,0.9,1.5,0)$ & 0.0951&0.0968&0.0967&0.0964&0.0969&0.0960\\
$(0,0.99,1.5,0)$ & 0.1001&0.0973&0.0964&0.0938&0.0925&0.0928\\
$(0,1,1.5,0)$ & 0.0995&0.0993&0.0994&0.0983&0.00991&0.0979\\
$(0,0.9,0.5,0)$ & 0.1008&0.1004&0.0999&0.0978&0.0982&0.0977\\
$(0,0.99,0.5,0)$ & 0.0991&0.0980&0.0983&0.1033&0.1033&0.1032\\
$(0,1,0.5,0)$ & 0.0969&0.0963&0.0968&0.1033&0.1032&0.1032\\
$(-0.3,0.9,4,0)$ & 0.1687&0.1675&0.1657&0.1435&0.1436&0.1443\\
$(-0.3,0.99,4,0)$ & 0.4747&0.4658&0.4617&0.2669&0.2664&0.2658\\
$(-0.3,1,4,0)$ & 0.6717&0.6669&0.6623&0.3187&0.3173&0.3163\\
$(-0.1,0.9,1.5,0)$ & 0.2505&0.2343&0.2290&0.1994&0.1922&0.1915\\
$(-0.1,0.99,1.5,0)$ & 0.7035&0.6879&0.6850&0.5203&0.5124&0.5112\\
$(-0.1,1,1.5,0)$ & 0.8475&0.8401&0.8378&0.6384&0.6324&0.6308\\
$(-0.002,0.9,0.5,0)$ & 0.8341&0.8306&0.8304&0.8088&0.8079&0.8072\\
$(-0.002,0.99,0.5,0)$ & 0.9728&0.9723&0.9722&0.9547&0.9545&0.9545\\
$(-0.002,1,0.5,0)$ & 0.9885&0.9888&0.9888&0.9683&0.9682&0.9682\\
\hline
\end{tabular*}
\end{table}

In Tables \ref{t3}~and~\ref{t4}, we report the powers for these tests. We choose
$a=-0.3$, $-$0.1, $-$0.002 for $\nu=4$, $1.5$, $0.5$, respectively.
From these two tables, we observe that the proposed empirical
likelihood method with known $\alpha$ is much more powerful than
the one with unknown $\alpha$ especially for the case of $\nu=4$. When
the normal approximation method produces a consistent size, that is,
the case of $(\phi,\nu)=(0.9,4)$, it is more powerful than the proposed
empirical likelihood methods in both (\ref{emplik0}) and (\ref{emlik}).
When the Bonferroni \mbox{$Q$-}test with known $\alpha$ and $\theta$ has a
consistent size, that is, the case of $(\phi,\nu,b_1)=(1,4,0)$, it is
more powerful than the proposed empirical likelihood method in (\ref
{emlik}), but less powerful than the empirical likelihood method in
(\ref{emplik0}).

\begin{table}[b]
\tabcolsep=0pt
\caption{Confidence intervals for the monthly CRSP value-weighted
index are computed for the proposed empirical likelihood method}\label{t6}
\begin{tabular*}{\tablewidth}{@{\extracolsep{\fill}}@{}lccccc@{}}
\hline
\textbf{CRSP series} & \textbf{Variable}&$\bolds{\hat\beta_{\mathrm{LSE}}}$&$\bolds{\hat\sigma_{V}/\hat\sigma_U}$&$\bolds{I_{0.9}}$&$\bolds{I_{0.95}}$\\
\hline
1926--2002&$d$--$p$&0.0083&1.0367&$[-0.0042, 0.0231]$&$[-0.0068, 0.0259]$\\
1926--2002&$e$--$p$&0.0129&1.0428&\phantom{$-$}$[0.0034, 0.0317]$&\phantom{$-$}$[0.0008, 0.0346]$\\
1926--1994&$d$--$p$&0.0123&1.0342&$[-0.0134, 0.0297]$&$[-0.0175, 0.0342]$\\
1926--1994&$e$--$p$&0.0211&1.0373&$[-0.0059, 0.0401]$&$[-0.0102, 0.0449]$\\
1952--2002&$d$--$p$&0.0116&1.0324&$[-0.0105, 0.0181]$&$[-0.0133, 0.0208]$\\
1952--2002&$e$--$p$&0.0088&1.0117&$[-0.0134, 0.0118]$&$[-0.0159, 0.0142]$\\
\hline
\end{tabular*}
\end{table}

It is easy to verify that Theorems \ref{th1} and \ref{th2} still hold when $Z_t(\beta)$
in (\ref{emplik0}) and $\tilde Z_t(\beta)$ in (\ref{emlik}) are
replaced by $Z_t(\beta)=(Y_t-\beta X_{t-1})X_{t-1}/w(X_{t-1})$ and
$\tilde Z_t(\beta)=(\tilde Y_t-\beta\tilde X_{t-1})\tilde
X_{t-1}/w(\tilde X_{t-1})$, respectively, for some weight function
$w(t)$ satisfying that $w(t)/t$ converges to a positive constant as
$t\to\infty$. A theoretical optimal weight function will be chosen to
minimize the coverage probability error. Without doubt, it is
impossible to obtain such an optimal one. Here we\vspace*{1pt} consider the class
$w(t)=(1+|t|^{h})^{1/h}$ for some $h>0$. Under the same setup as above,
we compute the size and power for the proposed empirical likelihood
methods for $h=1$, $2$, $4$. From Table~\ref{t5}, we observe that the methods
are not quite sensitive to the choice of $h$ especially when $X_t$ has
an infinite variance.

To summarize the simulation results, we find the reliable evidence that
the proposed empirical likelihood method in (\ref{emlik}) can deliver
an accurate size and a nontrivial power regardless of the predicting
variable being stationary or near-integrated, or having an infinite variance.

\section{Predictability of monthly CRSP value-weighted index}\label{sec4}

A frequently asked question in financial econometrics is whether asset
returns can be predicted by some macroeconomic data such as the
dividend-price ratio and the earnings-price ratio as well as other
state variables like interest rates.
In this section we apply the empirical likelihood method in (\ref
{emlik}) to re-visit the data set analyzed by \citet{r5}.
More specifically, the predictable variable $Y_t$ is the monthly CRSP
value-weighted index data (1926:12--2002:12) from the Center for
Research in Security Prices,
and the predicting variable $X_t$ is either the log dividend-price
ratio (ldp) or the log earnings-price ratio (lep).
The dividend-price ratio is computed as dividends over the past year
divided by the current price, and the earnings-price ratio is computed
as a moving average of earnings over the past
ten years divided by the current price. There are 913 observations in
total. The detailed description of this data set can be found in
\citet{r5}. Similar to \citet{r5}, we
consider three time periods as \mbox{1926:12--2002:12}, 1926:12--1994:12 and
1952:12--2002:12. The main purpose of revisiting this particular data
set is to argue that the proposed methodology in this paper can provide
more accurate
statistical inference than that in \citet{r5}.

Based on the above data set and model (\ref{Mod2}) with $p=0$, \citet{r5} calculated the Bonferroni \mbox{$Q$-}test for $\tilde\beta
=\beta\sigma_V/\sigma_U$ rather than $\beta$ by
simply scaling the test by $\hat\sigma_V/\hat\sigma_U$, where $\sigma
_V, \hat\sigma_V$ and $\sigma_U, \hat\sigma_U$ denote the standard
deviation and estimated standard deviation of $V_t$ and $U_t$ in (\ref
{Model}), respectively.
Hence, the results in Table~5 of \citet{r5} ignored the
effect of the plug-in estimators $\hat\sigma_U$ and $\hat\sigma_V$. It
is natural to conjecture that such an effect should result in wider intervals
for $\beta$ than those reported in Table~5 of \citet{r5}.
Moreover, due to the complicated implementation and too simplified
theoretical derivations in \citet{r5},
one may question the reliability of the empirical findings in \citet{r5}.
Here, we employ the proposed empirical likelihood method in (\ref
{emlik}) to compute intervals for $\beta$ rather than $\tilde\beta$.
Since the new method works for all cases with sound theory and is easy
to implement, we believe that the analysis under the new method is more
robust and reliable.

Table~\ref{t6} reports confidence intervals with levels $0.90$ in the fifth
column and $0.95$ in the last column for the monthly CRSP
value-weighted index with periods \mbox{1926--2002}, 1926--1994 and 1952--2002
as in Table~5 of \citet{r5}.
It is not surprising to observe from Table~\ref{t6} that the new intervals
are indeed wider than those reported in Table~5 of \citet{r5} because, as argued earlier, \citet{r5} ignored the
effect of plug-in estimators.
Similar to \citet{r5}, the null hypothesis of no
predictability ($H_0\dvtx  \beta=0$) is not rejected by the new method for
the log dividend-price ratio for all three time periods and for the log
earnings-price ratio in the subsample 1952--2002. Also, the null
hypothesis of no predictability is rejected by the new method for the
log earnings-price ratio for the full sample 1926--2002 at both levels
$90\%$ and $95\%$.
However, interestingly, the null hypothesis of no predictability is
not rejected by the proposed new method for the log earnings-price
ratio in the subsample 1926--1994, while it is rejected by \citet{r5}.
Indeed, our finding for this subsample are similar to the conclusion
in \citet{r3} for the period 1930:12--1990:12. That is, the
asset return is not predictable in the subsample through the
early 1990s.
The source of this difference between our finding and the result in
\citet{r5} can be
explained by the following arguments.
For this subsample, the confidence interval for $\phi$ [see Table~4 in
\citet{r5}] is
$[0.970, 0.997]$ and it does not cover $\phi=1$ so that $X_t$ might be
stationary and is a less persistent series.
As indicated earlier, the Bonferroni \mbox{$Q$-}test may not perform well
when $X_t$ is stationary or nearly integrated.\looseness=1

\section{Conclusion}\label{sec5}

Researchers have constantly asked whether stock returns can be
predicted by macroeconomic data. However, macroeconomic data may
exhibit nonstationarity and heavy tails.
Therefore, it is important to have a unified method to test
predictability in regressions without
distinguishing whether the predicting variable is stationary or
nonstationary or has an infinite variance.

In this paper, we study a predictive regression model which has an
ability to include the regressors to be a stationary or nonstationary
(integrated/nearly integrated) process and/or has an infinite variance and
allows the so-called two innovations to be correlated.
We propose novel empirical likelihood methods based on some weighted
score equation to construct a confidence for the coefficient or to test
the predictability.
We show that Wilks' theorem holds for the proposed empirical likelihood
methods regardless of the predicting variable being stationary, or
nonstationary or having an infinite variance.
The proposed new methods are easy to implement without any ad hoc
method such as the bootstrap method for obtaining critical values.
Therefore, the proposed new methods provide more robust findings than
other existing methods in the literature of predictive regressions and
have wide applications in financial econometrics.

\section{Proofs}\label{sec6}

We only prove Theorem~\ref{th2} since the proof of Theorem~\ref{th1} is easier.

\begin{pf*}{Proof of Theorem~\ref{th2}}
Put $\tilde V_j=V_j-V_{j+m}$ and let $\mathcal{F}_t$ denote the $\sigma
$-field generated by $\{(\tilde U_{s}, \tilde V_{s})\dvtx  1\le s\leq t\}\cup
\{V_s\dvtx  s\le0\}$.
Write $B(L)=\prod_{j=1}^p(1-\tilde b_jL)$. Then we have
$B^{-1}(L)=\prod_{j=1}^p(1-\tilde b_jL)^{-1}=\sum_{k=0}^{\infty}a_kL^k$ and
\[
e_t=\sum_{k=0}^{\infty}a_kV_{t-k}=
\sum_{k=0}^{t-1}a_kV_{t-k}+
\sum_{k=t}^{\infty}a_kV_{t-k}.
\]
Note that
%
\begin{equation}
\label{pf2-0} |a_k|\le k^p \Bigl(\max
_{1\le i\le p}|\tilde b_i| \Bigr)^k\quad
\mbox{and}\quad\max_{1\le i\le p}|\tilde b_i|<1.
\end{equation}
Put $e_{t,1}=\sum_{k=0}^{t-1}a_k\tilde V_{t-k}+\sum_{k=t}^{\infty
}a_kV_{t-k}-\sum_{k=t+m}^{\infty}a_kV_{t+m-k}$ and $e_{t,2}=\break -\sum_{k=t}^{t+m-1}a_kV_{t+m-k}$. Then we have
%
\begin{equation}
\label{pf2-1} e_t-e_{t+m}=e_{t,1}+e_{t,2}
\qquad\mbox{for } t=1,\ldots,m.
\end{equation}
Write
%
\begin{equation}
\label{pf2-2} X_t=\frac{1-\phi^t}{1-\phi}\theta+\sum
_{j=1}^t\phi^{t-j}e_j+
\phi^tX_0
\end{equation}
and
%
\begin{eqnarray}
\label{pf2-3} X_{t+m}&=&\frac{1-\phi^{t+m}}{1-\phi}\theta+\sum
_{j=1}^{t+m}\phi ^{t+m-j}e_j+
\phi^{t+m}X_0\nonumber
\\
&=&\frac{1-\phi^t}{1-\phi}\theta+\sum_{j=1}^t
\phi^{t-j}e_{j+m}+\phi^tX_0+
\frac{\phi^t-\phi^{t+m}}{1-\phi}\theta
\\
&&{}+\sum_{j=1}^m\phi
^{t+m-j}e_j+ \bigl(\phi^{t+m}-\phi^t \bigr)X_0.
\nonumber
\end{eqnarray}
Put
$W_{t,1}=\sum_{j=1}^t\phi^{t-j}e_{j,1}$ and
\[
W_{t,2}=\sum_{j=1}^t
\phi^{t-j}e_{j,2}-\frac{\phi^t-\phi^{t+m}}{1-\phi
}\theta-\sum
_{j=1}^m\phi^{t+m-j}e_j- \bigl(
\phi^{t+m}-\phi^t \bigr)X_0.
\]
Then, it follows from (\ref{pf2-1})--(\ref{pf2-3}) that
%
\begin{equation}
\label{pf2-4} \tilde X_t=W_{t,1}+W_{t,2}\qquad
\mbox{for } t=1,\ldots,m.
\end{equation}
When $|\phi|<1$, it follows from (\ref{pf2-0}) that as $n\to\infty$
%
\begin{eqnarray}\label{eq4}
\qquad \frac{1}{m}\sum_{t=1}^m
\frac{W_{t-1,1}^2}{1+W_{t-1,1}^2} &=&\frac{1}m\sum_{t=1}^m
\Biggl(\sum_{j=1}^t\phi^{t-j}e_{j,1}
\Biggr)^2\bigg/ \Biggl\{1+ \Biggl(\sum_{j=1}^t
\phi^{t-j}e_{j,1} \Biggr)^2 \Biggr\}
\nonumber
\\
&=&\lim_{t\to\infty} E\frac{(\sum_{j=1}^{t}\phi
^{t-j}e_{j,1})^2}{1+(\sum_{j=1}^{t}\phi^{t-j}
e_{j,1})^2}+o_p(1)
\\
&:=&\sigma^2_0+o_p(1). \nonumber
\end{eqnarray}
When $\phi=1-\gamma_\phi/n$ for some constant $\gamma_\phi\ge0$, we have
%
\begin{equation}
\label{pf2-5} |W_{t,1}|\overset{p} {\to}\infty,\qquad|W_{t,1}|=O_p
\bigl(t^{1/\alpha^\ast
} \bigr)\quad\mbox{and}\quad\frac{|W_{t,1}|}{t^{1/\alpha^\ast-\delta
_0}}\overset{p} {
\to}\infty
\end{equation}
for any $\delta_0>0$
as $t\to\infty$ by using (\ref{pf2-0}) and the fact that the
distribution of $V_t$ lies in the domain of attraction of a stable law
with index $\alpha^\ast$. Hence,
\[
\frac{W_{t-1,1}^2}{1+W_{t-1,1}^2}\overset{p} {\to}1\qquad\mbox{as } t\to \infty,
\]
that is,
%
\begin{equation}
\frac{1}{m}\sum_{t=1}^m
\frac{W_{t-1,1}^2}{1+W_{t-1,1}^2} \stackrel{p} {\to}1\qquad\mbox{as } n\to\infty.
\label{eq5}
\end{equation}
By (\ref{eq4}) and (\ref{eq5}), we have as $n\to\infty$
\begin{eqnarray*}
&& \frac{1}{m}\sum_{t=1}^mE \biggl(\frac{\tilde
U_t^2W_{t-1,1}^2}{1+W_{t-1,1}^2}\bigg| \mathcal{F}_{t-1} \biggr)
\\
&&\qquad =
2EU_1^2\frac{1}{m}\sum
_{t=1}^m\frac{W_{t-1,1}^2}{1+W_{t-1,1}^2}
\\
&&\qquad  \stackrel{p} {\to} \cases{2EU_1^2
\sigma_0^2,&\quad if $|\phi|<1$,
\vspace*{4pt}\cr
2EU_1^2,&
\quad if $\phi=1-\gamma_\phi/n$.}
\end{eqnarray*}
Similarly, for any $c>0$,
\begin{eqnarray*}
&& \frac{1}{m}\sum_{t=1}^mE \biggl(
\frac{\tilde
U_t^2W_{t-1,1}^2}{1+W_{t-1,1}^2}I \biggl(\frac{\tilde U_t^2W_{t-1,1}^2} {
1+W_{t-1,1}^2}>c^2m \biggr)\bigg|
\mathcal{F}_{t-1} \biggr)
\\
&&\qquad \le\frac{1}{(c\sqrt m)^q}\frac{1}{m}\sum_{t=1}^mE
\biggl( \biggl\llvert \frac
{\tilde U_tW_{t-1,1}}{\sqrt{1+W_{t-1,1}^2}} \biggr\rrvert ^{2+q}\bigg|
\mathcal{F}_{t-1} \biggr)
\\
&&\qquad =\frac{E|\tilde U_1|^{2+q}}{(c\sqrt m)^q}\frac{1}{m}\sum_{t=1}^m
\frac
{|W_{t-1,1}|^{2+q}}{(1+W_{t-1,1}^2)^{(2+q)/2}}
\\
&&\qquad \overset{d} {\to} 0\qquad\mbox{as } n\to\infty.
\end{eqnarray*}
By Corollary~3.1 of \citet{r12}, we have as $n\to\infty$
%
\begin{equation}
\label{eq6} \qquad \frac{1}{\sqrt m} \sum_{t=1}^m
\frac{\tilde U_tW_{t-1,1}}{\sqrt {1+W_{t-1,1}^2}}\stackrel{d} {\to} \cases{N \bigl(0,2E
\bigl(U_1^2 \bigr)\sigma^2_0
\bigr), &\quad if $|\phi|<1$,
\vspace*{4pt}\cr
N \bigl(0,2EU_1^2
\bigr),& \quad if $\phi=1-\gamma_\phi/n$.}
\end{equation}
Using (\ref{pf2-0}) and the fact that the distribution of $V_t$ lies in
the domain of attraction of a stable law with index $\alpha^\ast$, it
is easy to check that
%
\begin{equation}
\label{pf2-8} \bigl|\phi^{-t}W_{t,2}\bigr|=\cases{O_p(1),&
\quad if $|\phi|<1$,
\vspace*{2pt}\cr
O_p \bigl(m^{1/\alpha^\ast} \bigr),&\quad if $
\phi=1-\gamma_\phi/n$}
\end{equation}
and
%
\begin{equation}
\label{pf2-9} \frac{|W_{t,2}|}{t^{1/\alpha^\ast-\delta_0}}\overset{p} {\to}\infty
\end{equation}
for any $\delta_0>0$ as $t\le m$ goes to infinity. Hence, by (\ref
{pf2-5}), (\ref{pf2-8}) and (\ref{pf2-9}), we have
%
\begin{eqnarray}\label{eq7}
&& \frac{1}{\sqrt m}\sum_{t=1}^m \biggl(
\frac{\tilde U_t\tilde
X_{t-1}}{\sqrt{1+\tilde X_{t-1}^2}} -\frac{\tilde U_t W_{t-1,1}}{\sqrt{1+ W_{t-1,1}^2}} \biggr)
\nonumber
\\
&&\qquad =-\frac{1}{\sqrt m}\sum_{t=1}^m\tilde
U_t\frac{1}{\{1+(a_{t-1}\tilde
X_{t-1}+(1-a_{t-1})W_{t-1,1})^2\}^{3/2}}W_{t-1,2}
\nonumber
\\
&&\qquad =-\frac{1}{\sqrt m}\sum_{t=1}^m\tilde
U_t\frac{1}{\{
1+(W_{t-1,1}-a_{t-1}W_{t-1,2})^2\}^{3/2}}W_{t-1,2}
\nonumber\\[-8pt]\\[-8pt]
&&\qquad =\cases{ \displaystyle O_p \Biggl(\frac{1}{\sqrt m}\sum
_{t=1}^m\phi^t|\tilde U_t|
\Biggr),&\quad if $|\phi|<1$,
\vspace*{5pt}\cr
\displaystyle O_p \Biggl(
\frac{1}{\sqrt{m}}\sum_{t=1}^m|\tilde
U_t|\frac
{m^{1/\alpha^\ast+\delta_0}}{t^{3(1/\alpha^\ast-\delta_0)}} \Biggr),& \quad if $\phi=1-
\gamma_\phi/n$}\nonumber
\\
&&\qquad =o_p(1)\qquad\mbox{as } n\to\infty,\nonumber
\end{eqnarray}
where $a_{t-1}\in[0, 1]$ may depend on $\tilde X_{t-1}$ and
$W_{t-1,1}$, and $\delta_0>0$ is small enough. It follows from (\ref
{eq6}) and (\ref{eq7}) that
\[
\frac{1}{\sqrt m}\sum_{t=1}^m\tilde
Z_t(\beta_0)\overset{d} {\to}\cases{ N \bigl(0,
2EU_1^2\sigma_0^2 \bigr),&\quad
if $|\phi|<1$,
\vspace*{4pt}\cr
N \bigl(0, 2EU_1^2 \bigr),&\quad if $
\phi=1-\gamma_\phi/n$}
\]
as $n\to\infty$.
Similarly, we can show that
\[
\frac{1}m\sum_{t=1}^m\tilde
Z_t^2(\beta_0)\overset{p} {\to}\cases{
2EU_1^2\sigma_0^2,&\quad if $|
\phi|<1$,
\vspace*{4pt}\cr
2EU_1^2,&\quad if $\phi=1-
\gamma_\phi/n$}
\]
as $n\to\infty$.
The rest follows from the standard arguments in the proof of the
empirical likelihood method [see Chapter~11 of \citet{r19}].\
\end{pf*}

\section*{Acknowledgments}
We thank the Editor, Professor Susan Paddock, an Associate Editor and
two reviewers for helpful comments.



%

\printaddresses

\end{document}